# Feedback and harmonic locking of slot-type optomechanical oscillators to external low-noise reference clocks

Jiangjun Zheng,[1,*] Ying Li,[1] Noam Goldberg,[1] Mickey McDonald,[2] Xingsheng Luan,[1] Archita Hati,[3] Ming Lu,[4] Stefan Strauf,[5] Tanya Zelevinsky,[2] David A. Howe,[3] and Chee Wei Wong,[1,*]

[1] Optical Nanostructures Laboratory, Columbia University, New York, NY 10027, USA

[2] Department of Physics, Columbia University, New York, NY 100027-5255, USA

[3] National Institute of Standards and Technology, Boulder, CO 80303, USA

[4] Center for Functional Nanomaterials, Brookhaven National Laboratory, Upton, NY 11973, USA

[5] Department of Physics and Engineering Physics, Stevens Institute of Technology, Hoboken, NJ 07030, USA

**Abstract:** We demonstrate feedback and harmonic locking of chip-scale slot-type optomechanical oscillators to external low-noise reference clocks, with suppressed timing jitter by three orders of magnitude. The feedback and compensation techniques significantly reduce the close-to-carrier phase noise, especially within the locking bandwidth for the integral root-mean-square timing jitter. Harmonic locking via high-order carrier signals is also demonstrated with similar phase noise and integrated root-mean-square timing jitter reduction. The chip-scale optomechanical oscillators are tunable over an 80-kHz range by tracking the reference clock, with potential applications in tunable radio-frequency photonics platforms.

[*] Electronic addresses: jz2356@columbia.edu and cww2104@columbia.edu

In recent years, there have been much research progresses in the field of mesoscopic cavity optomechanics [1]. In the optomechanical (OM) cavities, the optical and mechanical modes are well coupled: the strong localized optical field induces scattering or gradient optical force that perturbs the mechanical structure, meanwhile the mechanical motion changes the phase of the intra-cavity optical field producing a reciprocal shift of the optical resonance. The frequency and damping rate of the mechanical modes are optically tunable due to the "dynamic backaction", by pumping the OM cavity with a continuous-wave laser [1−10]. It is interesting that when the intrinsic mechanical damping is overcome by the OM amplification, parametric oscillations can be observed [11-13], i.e., the OM cavity becomes an oscillator with self-sustained mechanical oscillations recorded by the periodic optical transmission along with linewidth reduction. For potential applications in radio-frequency (RF) photonics and low phase noise oscillators [14-19], fundamental characteristics of the OM oscillators (OMOs) were demonstrated with micrometer-scale high quality factor ($Q$) microtoroids [14,15].

Recent measurements [8,14] with tens of MHz OMOs shows that they are potential frequency sources with high long-term stability. However, compared to typical temperature compensated crystal oscillators (TXCOs), the phase noise of the OMOs is over tens of dBc/Hz higher, especially at low frequency offset. This excess phase noise [20, 21] can be suppressed in a variety of different ways such as electronic active noise cancellation for improving the long-term frequency stability of OMOs [22] or distributed oscillators [23]. Injection locking in microtoroid OMOs [24] or in photonic crystal OMOs by our team has been observed with reduced phase fluctuations via direct amplitude modulation of the optical pump with a single tone RF signal. In this Letter, we demonstrate the locking of two-dimensional slot-type photonic crystal OMOs [13] to low-noise external reference clocks through both feedback locking and higher harmonic locking schemes. The feedback error signal is obtained by comparing the chip-scale OMO with external reference signals. Once locked, the slot-type OMO tracks the reference up to an 80 kHz frequency range, larger than previously reported in microtoroids due to the large optical spring effect in the slot-type OMOs. Significant noise reduction is achieved at close-to-carrier offset frequencies, with illustrative more-than-40 dB reduction at 1 kHz offset. These locking schemes offer possibilities for chip-scale optomechanical oscillators in RF photonics.

The experimental scheme is shown in Fig. 1(a). The light source is a high performance tunable laser (Santec TSL510 type C, $\lambda$ = 1500−1630 nm). Two polarization controllers are used to change the polarization of light incident on the electro-optic modulator (EOM) and the fiber taper, respectively. The fiber taper works as a microprobe by coupling light into and out of the optical cavity, a slot-type OMO [13] as shown in Fig. 1(b). This OMO is operated at about 65 MHz with intrinsic optical $Q$ of $4.2 \times 10^4$, large vacuum OM coupling rate of 782.6 kHz, and small effective mechanical mass of ~ 6.11 pg. It is worth to note that recently a method of using frequency calibration is presented in detail for accurate measurement of vacuum OM coupling rate [25]. With this method, the vacuum OM coupling rate of our device is determined to be 782.6 kHz. A fast detector monitors the optical intensity oscillations. With relative high input power, the radio-frequency (RF) spectrum of the optical transmission exhibits higher-order harmonics up to 1 GHz (~ 15th harmonic) which is limited by the detector bandwidth, due to the internal nonlinear OM transduction [13]. An electrical band-pass filter selects one specific harmonic, which is then mixed with the reference clock signal from a low-noise signal generator (SRS SG384). The reference with a high resolution of 1 $\mu$Hz is easily tuned to have almost identical frequency as the OMO, which is useful for easy initial locking. After mixing, the differential-frequency error signal is then fed into a tunable loop filter with PI$^2$D transfer function, with proportional (P), integral (I), and differential (D) feedback along with a second integral feedback (I). The final output signal is applied to the RF port of the EOM to modulate the input optical power. Measurements here are taken at room-temperature in air. While injection locking by direct modulation of the input optical power with frequency close to the OMO has also been observed, here our scheme employs a feedback loop and the error signal reflects the frequency deviation of the OMO from the reference.

The above feedback scheme is based on optical-induced mechanical frequency shift. The mechanical frequency of the OMO is tunable by changing the laser-cavity detuning and pump power. Fig. 1(c) shows an example RF power spectrum density (PSD) versus laser wavelength with a low input power of 20 $\mu$W by using a spectrum analyzer (Agilent E4440A). It is clear that the mechanical frequency at the blue side of the optical resonance is higher than that at the red side. By varying the input power at optimized detuning, a frequency tuning range larger than 800 kHz was demonstrated for our OMO. The frequency of an unlocked or unstabilized OMO can be easily affected by internal and external noise sources, with frequency instabilities and sizable root-mean-square (RMS) timing jitters which need to be reduced for practical RF photonic applications. The feedback scheme in Fig. 1(a) is well-suited for this purpose. Any frequency deviation from the reference will lead to a change in the error signal and thus a change of the controlled input power, under proper feedback loop settings for improved frequency stability. In our experiments, the feedback loop corner frequencies and gains are tuned sequentially to minimize the phase noise at the close-to-carrier offset frequencies. In this way, the frequency noise will be largely compensated within the locking bandwidth.

Fig. 2(a) shows the spectrum of the OMO before and after it is locked to the reference signal. A detector with 125 MHz bandwidth (New Focus Model 1811) is used for this fundamental mechanical frequency about 65 MHz. Compared to the unlocked OMO, the center frequency is locked and shifted to the reference frequency. The width of the center peak is much narrower, while the outer region remains similar in amplitude. In the measurements, the depth of center peak is proportional to the gain amplitude, and the separation of the side peaks is proportional to the locking bandwidth. This kind of wing-like structure indicates locking in the spectral domain, which is similar to those reported in laser locking demonstrations [26]. The time-domain indication of the locking is the error signal given by the loop-filter, as shown in Fig. 2(b). Before locking, the error signal is approximately sinusoidal with slightly different frequencies between the OMO and the reference. Its period is varying over time as the OMO frequency is slowly varying. Once locked, the error signal become noise-like with a near-zero average voltage. Its RMS amplitude is much smaller than that of the unlocked error signal. In our measurements, we observe that it is smallest within a locking bandwidth of 20 kHz. This locking bandwidth also ensures optimum phase noise performance and is used in the subsequent single sideband phase noise measurements. Furthermore, after locking, the OMO frequency is tracked down to the reference frequency, as shown in Fig. 2(c). The OMO is locked to the reference over an 80 kHz range by adjusting the reference frequency only, much larger than that with injection locking schemes in earlier implementations, and limited by the locking electronics bandwidth. If the laser wavelength is tuned to set a new free-running OMO frequency, the effective frequency tracking range will be expanded further. This large frequency tracking range is mainly attributed to the large spring effect of the slot-type OMO with a small effective mass (~ 6.11 pg) and a large vacuum OM coupling rate (782.6 kHz), for tunability in RF applications.

    The OM oscillation in the optical transmission contains many high-order harmonics. For the OMO here, the harmonics with frequency up to 1 GHz can be easily measured with a 1.3

mW input power [13]. The optical spring effect and the noise of OM oscillation are also reflected in these high-order harmonics. In principle, harmonic locking, i.e., locking by mixing higher-order harmonic signal and the reference signal is possible for our OMO. Actually, higher-order harmonics are preferred for many locking techniques, such as synchronization of a voltage controlled oscillator (VCO) to a master laser [23], because the shorter period of the RF signal has more phase sensitivity for locking. Here, the higher-order harmonic OMO signal is selected by the electrical filter and then mixed with the higher reference frequency. Fig. 3 illustrates the harmonic locking, such as with the 2nd-harmonic and 3rd-harmonic signals respectively. The reference frequencies are 130.055 MHz and 195.125 MHz respectively. A detector with 1 GHz bandwidth (New Focus Model 1611-AC) is used. The center peaks are resolution-limited with a measurement resolution bandwidth of 1 kHz, which indicates a reduced phase noise in the close-to-carrier offset frequency region.

Although the RF spectrum is able to show locking, phase noise measurements are necessary to further investigate the performance of the lock. Here, a phase noise analyzer (Symmetricom 5125A) at NIST is used for measurement, with a low noise 5 MHz quartz oscillator providing its reference signal for a low measurement noise floor. As shown in Fig. 4, the noise of the NIST measurement system with a 65 MHz input is over 20 dB lower than those of other measured signals. The noise floor is about -145 dBc/Hz at a 10-Hz offset and -160 dBc/Hz at 1-kHz offset, which ensures the measurements are sufficiently accurate. The NIST system is especially useful for low noise oscillators even in the close-to-carrier offset frequency region. After locking, the phase noise of the OMO is greatly suppressed, as shown by curves (4) − (6), which are for 1st-, 2nd- and 3rd-harmonic locking examples separately. The close-to-carrier phase noise with offset frequency less than 10 kHz is greatly suppressed. At 10-Hz offset, the phase noise is about −90 dBc/Hz, indicating a 90 dB reduction from the unlocked case, only about 10 dB higher than the reference clock. The phase noise at 100-Hz and 1-kHz offset are reduced by about 65 dB and 50 dB respectively. The spikes on the phase noise curves are quite similar in feature, which are inherited from the unlocked OMO. When unlocked, the OMO frequency is divided by 10 with corresponding reduction of phase noise by 20 dB, to be measurable by the Symmetricom system. When unlocked, the OMO is noisy with phase noise approaching 0 dBc/Hz at 10-Hz offset. The reference signal is also given by curve (2) in Fig. 4. Compared to the free-running OMO, the reference clock has a phase noise less than -100 dBc/Hz at 10-Hz offset. Its phase noise decreases to -140 dBc/Hz at 100-kHz offset, approximately over 40 dB better than the current free-running OMO.

While the phase noise level of the locked cases is close to the phase noise of reference within less than 1-kHz offset, for offset frequency more than 20-kHz the curves become similar and close to the free-running OMO. The small deviations of the curves beyond this offset frequency are due to different locking settings for each measurement, wherein slightly different optical detunings, reference frequencies, and loop gains are adjusted to minimize the measured phase noise. In this particular case the 2nd-harmonic locking is slightly better than the 1st- and 3rd- harmonic locking because, in this specific scenario, the electrical power in the 2nd-harmonic is slightly higher and with a different set of optimized locking parameters. After locking, the RMS timing jitter are greatly reduced, as listed in Table I. In the bandwidths of 10 − 100 Hz, 100 Hz− 1 kHz, and 1 kHz − 10 kHz, the integrated RMS timing jitters decreases by 8,400, 1,700, and 36.8 times respectively, suppressed down to 1.5 picosecond timing jitter over the 10 to 100 Hz range for the locked 65 MHz OMO.

In conclusion we have demonstrated locking of a chip-scale OMO to a low-noise external clock with harmonic and feedback locking. The noise-like error signal is obtained by comparing the frequencies of the OMO and the reference. It modulates the input power for compensating the frequency noises introduced by all internal and external noise sources. When locked and within the locking bandwidth, the phase noise of the stabilized on-chip OMO is greatly suppressed by locking different harmonic signals, especially up to nine orders of magnitude at close-to-carrier frequency offset. The integrated timing jitter is demonstrated to decrease significantly, suppressed down to 1.5 picosecond for the 65 MHz OMO. Additionally, the chip-based OMO remains locked and tunable over 80-kHz by changing the reference frequency. We believe the technique is interesting for locked and stabilized OMOs, in single units or arrays, for RF photonic applications. Future efforts include further reduction of the unlocked OMO phase noise and higher frequency oscillators, locking to low-noise Ti:Sapphire laser oscillators, and variant feedback schemes for distributed locking.

The authors thank discussions with Harish Krishnaswamy, Hong X. Tang, and Xingsheng Luan. This work is supported by Defense Advanced Research Projects Agency (DARPA) DSO with program manager Dr. J. R. Abo-Shaeer under contract number C11L10831. Device fabrication is carried out in part at the Center for Functional Nanomaterials, Brookhaven National Laboratory, which is supported by the U.S. Department of Energy, Office of Basic Energy Sciences, under Contract No. DE-AC02-98CH10886.

FIGURE

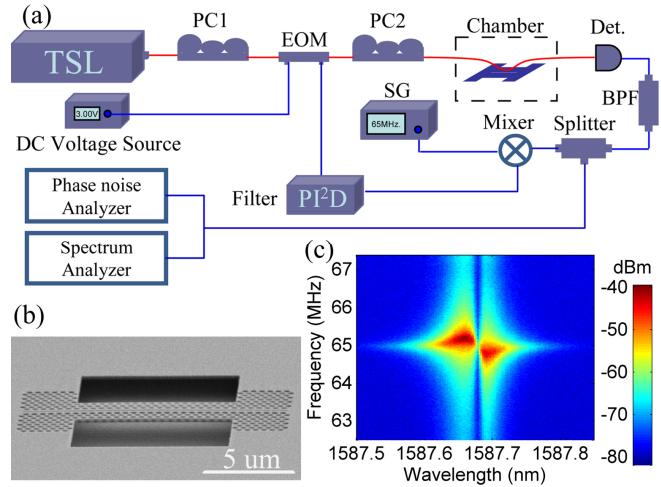

FIG. 1. (Color online) (a) Schematic diagram of the experimental setup used for feedback-locking of an air-slot type OMO. TSL indicates the tunable scanning laser (Santec TSL-510V, 1510 − 1630 nm). PC1 and PC2 indicate two polarization controllers. EOM indicate the intensity electro-optic modulator for intensity modulation. A DC voltage source is used to tune the bias voltage of the EOM. Det. indicates the fast detector that is used. BPF indicates a band pass filter for obtaining the selected harmonics. SG is a single tone RF signal generator (SRS SG384), which serves as the low phase-noise reference. The error signal from the mixer is then input to a tunable loop filter that provides PI$^2$D transfer function. Finally, the output signal is applied to the RF port of the EOM to vary the incident power. The RF OMO signal is coupled out by a power splitter and monitored by a phase noise analyzer (Symmetricom 5125A) and a spectrum analyzer (Agilent E4440A). (b) SEM image of the fabricated sample. (c) Example RF PSD by sweeping the laser wavelength from 1587.5 nm to 1587.85 nm with an input power of 20 $\mu$W.

FIGURE

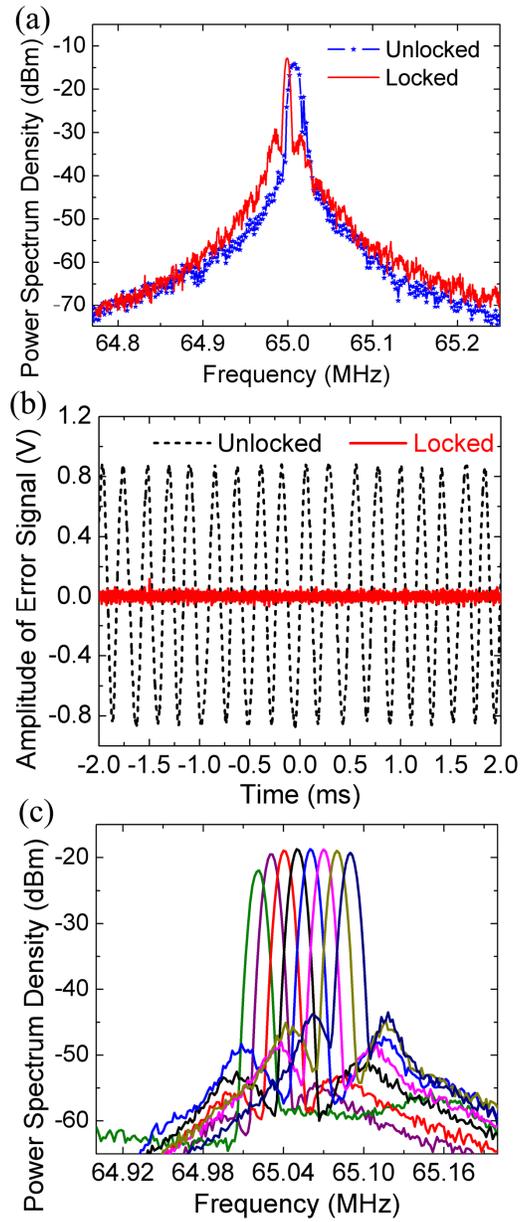

FIG. 2. (Color online) (a) Power spectrum of the OMO before (blue) and after (red) it is locked to the reference signal. (b) Typical error signal output from the mixer before and after the OMO is locked to the reference. (c) Power spectrum of the locked OMO tuned by the reference over an 80 kHz range.

FIGURE

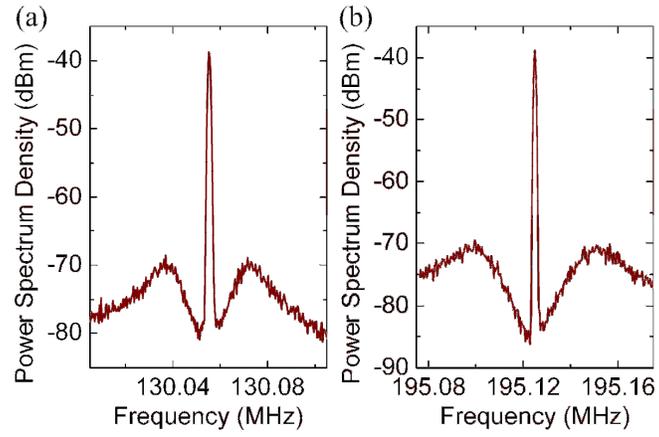

FIG. 3. (Color online) Power spectrum of the OMO locked to the reference by using 2nd-harmonic signal (a) and 3rd-harmonic signal (b). The resolution bandwidth is 1 kHz.

FIGURE

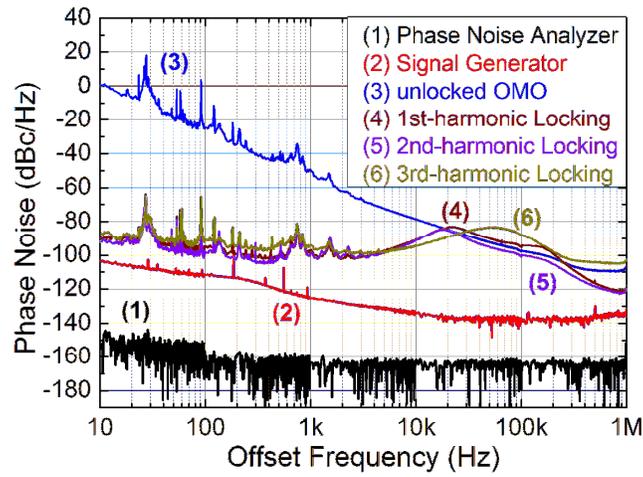

FIG. 4. (Color online) Measured phase noise spectral densities: (1) phase noise floor of the phase noise analyzer; (2) phase noise of the reference at 65 MHz; (3) phase noise of the free-running OMO; (4), (5) and (6) indicate the phase noise of 1st-harmonic locking, 2nd-harmonic locking, and 3rd-harmonic locking respectively.

TABLE I.:

TABLE I. RMS timing jitter integrated over different bandwidths for the reference, free-running OMO and locked OMO by using different harmonic signals.

| Case \ Bandwidth[*] | 10 – 100 | 100 – 1k | 1k – 10k | 10k – 100k |
|---|---|---|---|---|
| **Signal Generator** | 0.12 | 0.13 | 0.10 | 0.14 |
| **unlocked OMO** | $1.26 \times 10^4$ | $3.41 \times 10^3$ | 210.45 | 36.68 |
| **1st-harmonic locking** | 1.72 | 2.10 | 5.65 | 39.01 |
| **2nd-harmonic locking** | 1.50 | 2.05 | 5.71 | 27.75 |
| **3rd-harmonic locking** | 3.09 | 3.21 | 4.81 | 56.19 |

[*]: Timing jitter unit: ps; Bandwidth unit: Hz